\documentclass[aps,onecolumn,showpacs]{revtex4}
 
\begin{document}
 %\tighten
 \bibliographystyle{apsrev}
 \def\half{{1\over 2}}
 \def \D {\mbox{D}}
 \def\curl {\mbox{curl}\,}
 \def \ep {\varepsilon}
 \def \lleq {\lower0.9ex\hbox{ $\buildrel < \over \sim$} ~}
 \def \ggeq {\lower0.9ex\hbox{ $\buildrel > \over \sim$} ~}
 \def\beq{\begin{equation}}
 \def\eeq{\end{equation}}
 \def\ber{\begin{eqnarray}}
 \def\eer{\end{eqnarray}}
 \def \apl {ApJ, }
 \def \aps {ApJS, }
 \def \pd {Phys. Rev. D, }
 \def \prl {Phys. Rev. Lett., }
 \def \pl {Phys. Lett., }
 \def \np {Nucl. Phys., }
 \def \l {\Lambda}

  \def\apj{{Astroph.\@ J.\ }}
  \def\mn{{Mon.\@ Not.\@ Roy.\@ Ast.\@ Soc.\ }}
  \def\asta{{Astron.\@ Astrophys.\ }}
  \def\aj{{Astron.\@ J.\ }}
  \def\prl{{Phys.\@ Rev.\@ Lett.\ }}
  \def\pd{{Phys.\@ Rev.\@ D\ }}
  \def\nucp{{Nucl.\@ Phys.\ }}
  \def\nat{{Nature\ }}
  \def\plb {{Phys.\@ Lett.\@ B\ }}
  \def \jetpl {JETP Lett.\ }

\title{Manybody treatment of white dwarf and neutron stars on the brane}
\author{Mofazzal Azam$^1$ and \ M. Sami$^2$}
\address{$^1$Theoretical Physics Division, Bhabha Atomic Research Centre\\
Mumbai, India\\
$^2$Inter-University Centre for Astronomy and Astrophysics\\
Pune, India}
\begin{abstract}
Brane-World models suggest modification of Newton's
law of gravity on the 3-brane at submillimeter
scales. The brane-world induced corrections
are in higher powers of inverse distance and appear as additional
terms with the Newtonian potential. The average inter-particle
distance in white dwarf and neutron stars are $10^{-10}~cms$ and
$10^{-13}~cms$ respectively, and therefore, the effect of submillimeter
corrections needs to be investigated.
We show, by carrying out simple manybody
calculations, that the mass and mass-radius relationship of
the white dwarf and neutron stars are not effected by submillimeter
corrections.
However, our analysis shows that the correction terms
in the effective theory give rise to force akin to surface tension
in normal liquids.

\end{abstract}
\maketitle
\section{INTRODUCTION}
String theory is considered as a strong candidate for the unified
theory of the standard model and gravity. However, string theory is a
higher dimensional theory, and therefore, it is  necessary to find a
mechanism to get rid of the extra dimensions with some imprint left on the
four dimensional physical world. But such effects must be consistent
with the existing experimental and observational data.
The brane-world scenario proposed by Randall and Sundrum\cite{ran,ran1}
for describing a low dimensional
physical theory starting from a higher dimensional theory has attracted
lot of attention recently\cite{othors}. The model has triggered a large
amount of work in cosmology and astrophysics\cite{applications}.

The brane-world scenario leads to modification of gravity
at submillimeter distance scales.
The brane corrections are in higher powers of inverse distance
and appear as additional terms in the Newtonian potential. In the
Randall-Sundrum scenario, the modification suggested for the Newtonian
potential is given   by,
\begin{eqnarray}
\Phi(r)= \frac{GM}{r}(1+\frac{2l^2}{3r^2})
\end{eqnarray}
where $l$ is the curvature scale of 5-dimensional
anti-deSitter space time, and
$G$ is the usual Newton's constant of gravity, $M$ is the mass, $r$ is the
distance in 3-space. In the setting above, $l$ has the dimension of length
in 3-space, in other words, it is measured in centimeters in standard CGS
units. Henceforth, we will take $l_{s}^{2}=\frac{2}{3}l^2$

Several groups have estimated bounds on the distance scales at which
such corrections may be allowed without violating the existing experimental
and observational data \cite{dam}.
As mentioned above, there have also been some works dealing with
the astrophysical consequences of
such modifications of gravity \cite{roy1,roy2,wise}.
The average inter-particle
distance in white dwarf and neutron stars are $10^{-10}~cms$ and
$10^{-13}~cms$ respectively, and therefore, the effect of submillimeter
corrections needs to be investigated. Wiseman \cite{wise} has carried
out a detailed numerical study of the neutron stars on the brane and
claimed that the normal parameters of the neutron stars such as
mass and radius are not effected by brane world corrections.
In this paper, we carry out simple manybody analytical calculations
using the brane modified gravitational potential.
Our approach is based on manybody theory in
semi-relativistic quantum mechanics \cite{lieb,teuk}.
%At first, we briefly sketch the derivation of upper bound for
%the mass and the radius  of the white dwarf
%and neutron stars without any modification of
%the Newtonian gravity. We point out the problem we encounter
%if we trivially extend this method for the analysis of these
%stars on the brane. After this, we carry out a heuristic analysis of
%mass and radius of the compact stars on the brane and show that
%this type of analysis does not lead to sensible results.
At first, we briefly summerise the nature
of the potential and the force that emerges in the manybody context
due to the barneworld correction to gravity (the detail calculation
is presented in appendix-I).
We find that, for extended massive bodies,
the nature of the force and the potential due to brane correction terms
is very different from those due to Newtonian gravity. For example,
inside a  hollow spherical shell the potential due to Newtonian
gravity is constant, independent of location of the test particle.
Therefore, the force is zero inside the shell. We show
that this is not true in the presence of the brane correction term.
This result may have some relevance in context of
experimental observation of deviations from Newton's law of gravity.
We also show that the brane-world correction to gravity gives
rise to force akin to surface tension in normal liquids.
Following that we consider the effects due to the brane-world corrections
on the mass and the mass-radius relationship of white dwarf and neutron
stars. We find that the brane world corrections do not influence these
two parameters in any significant way.\\
In the appendix-I,
we provide detailed calculations of the potential, potential energy and
the force due to the brane correction term.
We also point out that many new physical features that emerge
from corrections to Newtonian gravity are generic
due to the deviation from the inverse square law of force
and does not depend on the particular form of the correction term.\\
In appendix-II, we analyse the nature of the extrema of various
energy functionals that appear in the text.

\section{Many-body Effects of Brane world Correction to Gravity}

We consider in this section the many-body effects of brane world
correction to gravity in a classical theory. Essentially, we carry
out analysis of the potential and force due to brane world correction
for spherically symmetric body of constant density and a fixed
radius. Details of calculations are given in appendix-I. Here we
provide a brief summary of these results. As already mentioned in the
introductory section, for extended massive bodies,
the nature of the force and the potential due to brane correction terms
is very different from those due to Newtonian gravity. These claims
are demonstrated latter in this section. \\
For a spherically symmetric ball of constant density
$~\rho~$ and radius $a$, the potential due to brane corrections
at a point $\vec{r}$ ($~r~>a~$, $r$ being the distance from centre of
the ball) outside the ball is given by,

\begin{eqnarray}
\Phi_B(r)=~-2\pi G~l_{s}^{2}~\rho~[~ln\frac{r+a}{r-a}~-~\frac{2a}{r}~]
                                                 \nonumber
\end{eqnarray}
The corresponding force is given by,
\begin{eqnarray}
F=2\pi Gl_{s}^{2}~\rho~[\frac{1}{r+a}-\frac{1}{r-a}+\frac{2a}{r^2}]
=-4\pi aGl_{s}^{2}~\rho~[\frac{1}{r^2-a^2}-\frac{1}{r^2}] \nonumber
\end{eqnarray}

When the point $\vec{r}$ is close to surface of the ball, both potential
and the force grows and becomes very large.Therefore, one needs to
introduce a parameter $\epsilon$ which characterises the smallest
inter-particle distance. At a distance $\epsilon$ from the surface of
the  ball, the force is given by,
\begin{eqnarray}
F=2\pi Gl_{s}^{2}\rho~[-\frac{1}{\epsilon}+\frac{2}{a}] \nonumber
\end{eqnarray}
This is clearly a skin effect. However, this force
is not as strong as one may
think. At a distance of $1~mm$ from the surface of a neutron star, this force
is weaker than the force of Newtonian gravity. The is because this force
depends on the density and not on the mass. However, this force
can be large close to the surface of a spherical grain of
very small radius and may even overtake the force due to
Newtonian gravity.

For a spherically symmetric shell of outer radius $R$ and inner
radius $b$, the potential at an interior point $\vec{r}$ ($r~<b~$) is
given by,
\begin{eqnarray}
\Phi'_B(r)=-2\pi Gl_{s}^{2}\rho~ ln\frac{R^2-r^2}{b^2-r^2} \nonumber
\end{eqnarray}
Force due to this potential, at distance $r$ from the centre is given by,
\begin{eqnarray}
F(r)=2\pi Gl_{s}^{2}\rho~[\frac{2r}{b^2-r^2}-\frac{2r}{R^2-r^2}] \nonumber
\end{eqnarray}
From the expression above it is obvious that $~~F(r)>0$.
This means that the force is directed in the direction in which
$r$ increases. In other words, it is directed towards the inner
boundary surface of the shell. At the centre of the shell,
$r=0$, and therefore,
the force is zero. But any small perturbation will move the test particle
away from the centre towards inner boundary surface.
This is a new element that emerges
from the brane-world correction to Newton's law of gravity. We know that the
force due to Newtonian gravity
is zero inside a spherically symmetric shell. If one can find the above
mentioned effect inside a shell, it will provide a very  clean test of
brane world corrections to Newtonian gravity. May be such experiments can be
carried out on board artificial satellites.
\par In this case, as in the case of spherical ball, there is skin effect.
The force of attraction close to the inner
boundary of the shell at distance, say, $\epsilon$ from the inner boundary
surface is given by,
\begin{eqnarray}
F=2\pi Gl_{s}^{2}\rho~[\frac{1}{\epsilon}-\frac{2b}{R^2-r^2}] \nonumber
\end{eqnarray}

Therefore, any interior layer of a spherical ball experiences two types
of forces:~
one is due to a sphere just bellow it and other is due to
the spherical shell outside.
These  two types of forces, individually,
give rise to a large forces on this layer
due to the skin effect.
Remarkably, these two forces act in opposite directions and the forces due
to skin effect are equal in magnitude and exactly cancel each other.
On the layer at $r$, let $F_1$ be the force due the inner sphere and $F_2$
be the force due to the outer shell.
\begin{eqnarray}
F_1=2\pi Gl_{s}^{2}\rho~[-\frac{1}{\epsilon}+\frac{2}{r}] \nonumber
\end{eqnarray}
\begin{eqnarray}
F_2=2\pi Gl_{s}^{2}\rho~[\frac{1}{\epsilon}-\frac{2r}{R^2-r^2}] \nonumber
\end{eqnarray}
Therefore, the net force at any point on the layer at a distance
$r~$ from the centre of the ball is given,
\begin{eqnarray}
F=F_1+F_2=2\pi Gl_{s}^{2}\rho~[\frac{2}{r}-\frac{2r}{R^2-r^2}] \nonumber
\end{eqnarray}
This result is valid for every layer except
the one at the outer most boundary.
For the outer most layer, there is an inner sphere but no outer shell.
There is no cancellation of the skin effect on this layer.
Therefore, it experiences a large force.This will give rise to effect
similar to surface tension in normal liquids.
\par The formula above also shows that the net force
due to the brane-world induced corrections at any point
on  layers at distance $r~$ where $r~<\frac{R}{\sqrt{2}}~$ is repulsive, and
on layers at a distance $r~$, $r~>~\frac{R}{\sqrt{2}}$ is attractive.
Within a  submillimeter central core of a neutron star there are
approximately $~10^{36}~$ neutrons arranged in
approximately  $~10^{12}~$ layers.
At any point on these layers the net repulsive force due to the brane-world
induced corrections overtake the attractive force due to Newtonian gravity.
Also note that any change in the radius of the star due to mass acreation
or mass loss leads to the rearrangement of the field of force in
the interior of the star. Therefore, the interior and the exterior
of the relativistic stars are intimately (although very weakly)
connected to each other in the presence of brane-world corrections.
\par The potential energy of a spherically symmetric body (of constant
density) of mass $M$ and radius $R$, due to the brane-world
induced corrections is given by,
\begin{eqnarray}
U= -\frac{3GM^2l_{s}^{2}}{R^3}ln\frac{R}{\epsilon}
=-\frac{3GN^2 m^2 l_{s}^{2}}{R^3}ln\frac{R}{\epsilon}   \nonumber
\end{eqnarray}
Here $N$ is the number of particles in the spherical ball of radius $R$,
$m$ is the mass of the elementary constituent particles,
In a classical theory, the parameter $\epsilon$
should be considered as the minimum inter-particle distance.
But in a quantum theory, as we will show in the latter section,
it is possible to fix its' value using Heisenberg's uncertainity
relation and  Pauli's exclusion principle.
%%%%%%%%%%%%%%%%%%%%%%%%%%%%%%%%%%%%%%%%%%%%%%%%%%%%%%%%%%%%%%%%%5

\section{Mass-Radius Relationship of the White Dwarf and Neutron Stars on
the Brane}

In this section, we show that the brane world corrections to Newton's
law of gravity does not influence the mass-radius relationship
of white dwarf and neutron statrs in any significant way.
Let us consider a system of large number of fermions contained
in some volume $V$. Let us also assume that the system is almost
a compact sphere with a large constant density such that fermions
are relativistic. We will consider both the white dwarf and neutron stars.
We do not discuss here the details of the nuclei present in
the white dwarf stars. We simply consider that,
in the white dwarf stars, there are two types of particles: the electrons and
the nucleons. For the nucleons, we have in mind the protons and the neutrons.
The over all charge neutrality of the star implies that number of electrons
and the protons should be equal. This suggests that for $N$ electrons in
the star there should be $2N$ nucleons ( $N$ protons and $N$ neutrons).
However, we will consider that the number of electrons is equal to the
number of nucleons which are essentially protons. Such a choice changes
the mass and radius of the white dwarf star by a small numerical
factor but the analysis is easily carried over to the case of neutron stars.
The zero-point kinetic energy in the white dwarfs is solely due to the
zero-point motion of the electrons but the gravitational potential
energy is  soley due to the gravitational interaction of the nucleons among
themselves. For the neutron stars the kinetic as well as potential energy is
generated by the neutrons alone. We assume that the mass of a nucleons is
equal to the mass of a neutron. We also consider a situation when the kinetic
energy due to temperature is much small compared to the zero point kinetic
energy and, therefore, can be neglected.\\

To make our derivation more transparent, at first, we derive the
mass-radius relationship without the brane correction terms. Let us consider
first the white dwarfs. We have a spherically symmetric region of
radius $R$ containing $N$ electrons and $N$ nucleons.
We assume the electrons to be relativistic; their total kinetic energy
is,
\begin{eqnarray}
K_N=~N~c~(p^2+m_{e}^{2}c^2)^{1/2} -N m_{e} c^2
\end{eqnarray}
Substituting the value of zero-point momentum, $p=\frac{\hbar N^{1/3}}{R}$
in the expression of the kinetic energy, we obtain the total energy as
\begin{eqnarray}
E(R)=
= N~c~(\frac{\hbar^2 N^{2/3}}{R^2}+m_{e}^{2}c^2)^{1/2} -N m_e c^2 \
 -\frac{3}{5}\times \frac{Gm_{n}^{2} N^2}{R}   \nonumber
\end{eqnarray}

The quantum mechanical variational ground state is obtained by demanding,
$\frac{dE(R)}{dR}~=0$.This gives,
\begin{eqnarray}
\frac{N^{1/3}\hbar}{(\hbar^2 N^{2/3}+m_{e}^{2}c^2 R^2)^{1/2}}
=\frac{3}{5}\times \frac{Gm_{n}^{2} N^{2/3}}{c \hbar}
\end{eqnarray}
Both the LHS and RHS of the equation above are positive. The LHS,
as is clear from equation above, is strictly less than one, and therefore,
the RHS is  strictly  less than one.In particular
\begin{eqnarray}
\frac{3}{5}\times \frac{Gm_{n}^{2} N^{2/3}}{c \hbar}~<~1    \nonumber
\end{eqnarray}
This provides an upper bound on
the number of particles and therefore, an upper bound on the mass
of the star. As before, the upper bound on the number of particles,
$N_{max}$, is given by,
\begin{eqnarray}
N_{max}^{2/3}=~\frac{5}{3}\times \frac{c \hbar} {Gm_{n}^{2}}
                                                              \nonumber
\end{eqnarray}
Therefore, as before the upper bound on the mass of the star is
$M_{max}=m_n N_{max}$.
Let us call the radius defined as,
\begin{eqnarray}
R_0~=~\frac{\hbar N^{1/3}}{m_e c}  \nonumber
\end{eqnarray}
Compton radius of a compact relativistic white dwarf. Let us also
introduce the notation,
\begin{eqnarray}
\alpha=~(N/N_{max})^{2/3} \nonumber
\end{eqnarray}
Using this, we can write Eq.(19) as
\begin{eqnarray}
\frac{1}{(1+~\frac{R^2}{R_{0}^2})^{1/2}}~=\alpha \nonumber
\end{eqnarray}
This can also be expressed as,
\begin{eqnarray}
R=R_0\sqrt{\frac{1}{\alpha^2}-1}
\end{eqnarray}

This is the mass-radius relationship of white dwarfs in the absence
of brane correction term. (It is shown in appendix-II, that this
solution corresponds to the minimum of the total energy, as a function
of $R$).
Note that $\alpha$, by definition, is strictly less than one. We can
get another bound on $\alpha$ as follows. The white dwarf star that
we are considering is
relativistic, and therefore, $p=\frac{\hbar N^{1/3}}{R}\geq m_e c$. This
implies that $R~\leq \frac{\hbar N^{1/3}}{m_e c}=R_0$. Taking $R\leq R_0$,
in the equation above, we find that $\alpha\geq \frac{1}{\sqrt{2}}$.
An order of magnitude calculation gives the mass of the white dwarf
between one to two solar masses and
the radius a few thousand kilometers.
Neutron star parameters are obtained simply by replacing the electron
mass by neutron mass. As an order of magnitude estimate,
the mass of the neutron stars turns out to be similar
to the white dwarf mass but the radius is just a few kilometers.
\par Let us now consider the effect of the brane-world induced
correction term on the parameters of white dwarf and the neutron stars.

\par The potential energy of spherically symmetric  ball due
to Randall-Sundrum brane corrections is given by,
\begin{eqnarray}
U_{RS}=~-\frac{3Gls^{2}m_{n}^{2}N^2}{R^3}\times ln\frac{R}{\epsilon}
                                                       \nonumber
\end{eqnarray}
Therefore, the total energy including brane correction is given by,
\begin{eqnarray}
E(R)= K_N~+~U_{N}(R)+~U_{RS}
= N~c~(p^2+m_{e}^{2}c^2)^{1/2} -Nm_{e}c^2
-\frac{3}{5}\times \frac{Gm_{n}^{2} N^2}{R}
-\frac{3Gls^{2}m_{n}^{2}N^2}{R^3}\times ln\frac{R}{\epsilon}
                                                         \nonumber
\end{eqnarray}
Substituting $p=\frac{\hbar N^{1/3}}{R}$, we obtain
\begin{eqnarray}
E(R)
= N~c~(\frac{\hbar^2 N^{2/3}}{R^2}+m_{e}^{2}c^2)^{1/2} -Nm_e c^2
-\frac{3}{5}\times \frac{Gm_{n}^{2} N^2}{R}
 -\frac{3Gls^{2}m_{n}^{2}N^2}{R^3}\times ln\frac{R}{\epsilon}
                                                   \nonumber
\end{eqnarray}

The quantum mechanical variational ground state is obtained as before,
%\begin{eqnarray}
%-\frac{c~ N^{5/3} \hbar^2}{R^3
%(\frac{\hbar^2 N^{2/3}}{R^2}+m_{e}^{2}c^2)^{1/2}}
%+\frac{3}{5}\times \frac{Gm_{n}^{2} N^2}{R^2}
%+\frac{9Gls^{2}m_{n}^{2}N^2}{R^4}\times ln\frac{R}{\epsilon}
%-\frac{3Gls^{2}m_{n}^{2}N^2}{R^4}=0      \nonumber
%\end{eqnarray}
%The last term, in the equation above, can be absorbed in the third term
%containing the logarithm by simply redefining the parameter $\epsilon$.
%Therefore, with the redefined $\epsilon$, the equation above takes the
%following form,
\begin{eqnarray}
\frac{N^{1/3}\hbar}{(\hbar^2 N^{2/3}+m_{e}^{2}c^2 R^2)^{1/2}}
=\frac{3}{5}\times \frac{Gm_{n}^{2} N^{2/3}}{c \hbar}
+\frac{9Gls^{2}m_{n}^{2}N^{2/3}}{R^2 c\hbar}\times ln\frac{R}{\epsilon}
\label{beq}
\end{eqnarray}
Note that $\epsilon$
was introduced as shortest inter layer distance in the spherical ball.
From physics point of view it can be identified with the shortest
inter-particle distance. For fermions, as per Pauli's exclusion
principle, it can not
be shorter than $R/N^{1/3}$.
We take $\epsilon=~R/N^{1/3}~$($~\ln\frac{R}{\epsilon}=\ln N^{1/3}~$).

Both the LHS and RHS of the equation above are positive. The LHS,
as is clear from equation above, is strictly less than one, and therefore,
the RHS is  strictly  less than one.In particular
\begin{eqnarray}
\frac{3}{5}\times \frac{Gm_{n}^{2} N^{2/3}}{c \hbar}~<~1 \nonumber
\end{eqnarray}
This provides an upper bound on
the number of particles $N$, and therefore, an upper bound on the mass
of the star, $M=m_n N$.
%%%%%%%%%%%%%%%%%%%%%%%%%%%%%%%%%%%%%%%%%
%Before, we come to the solution of the Eq.\ref{beq}, we need to have
%some idea about the value $\epsilon$ can take. 
%%%%%%%%%%%%%%%%%%%%%%%%%%%%%%%%%%%%%%%%%%%%%%
%Note that $\epsilon$
%was introduced as shortest inter layer distance in the spherical ball.
%From physics point of view it can be identified with the shortest
%inter-particle distance. For fermions, as per Pauli's exclusion
%principle, it can not
%be shorter than $R/N^{1/3}$.
%We take $\epsilon=~R/N^{1/3}~$($~\ln\frac{R}{\epsilon}=\ln N^{1/3}~$).
%%%%%%%%%%%%%%%%%%%%%%%%%%%%%%%%%%%%%%%%%%%%%%%%%%%%%%%%%%%%%%%
We will proceed in the same way as we did in the absence of the
brane correction terms. However, in this case we obtain a cubic
equation for the square of the radius. Before we
solve the equation, let us introduce the following notation,
\begin{eqnarray}
\beta = \frac{15ls^{2}}{R_{0}^2}\times ln\frac{R}{\epsilon}
      = \frac{15ls^{2}}{R_{0}^2}\times ln N^{1/3}
\end{eqnarray}
We introduce a new variable,
\begin{eqnarray}
Y=(\frac{R}{R_0})^2   \nonumber
\end{eqnarray}
The equation Eq.(\ref{beq}), in terms of the new variable can be written as,
\begin{eqnarray}
\frac{1}{(1+Y)^{1/2}}=\alpha +\frac{\alpha \beta}{Y} \nonumber
\end{eqnarray}
After some algebraic manipulations, we obtain
\begin{eqnarray}
Y^3+(2\beta+1-\frac{1}{\alpha^2}) Y^2+(2\beta+\beta^2) Y+\beta^2=0
\label{cubic}
\end{eqnarray}
Before we look for solution of this equation, let us have, an order
of magnitude, estimate of the values of parameters $\alpha$ and
$\beta$. They are both pure numbers. As mentioned earlier,
both for white dwarf and neutron stars, the range of $\alpha$ is given by
$\frac{1}{\sqrt{2}}\leq \alpha <1$. The parameter $\beta$, depends on
the Compton radius of the star, and therefore, its' value is different
for white dwarf and the neutron stars. Taking $N\approx 10^{57}$,
we find that for white dwarfs, $\beta\approx 10^{-16}$, and for neutron
stars, $\beta\approx 10^{-10}$.
%%%%%%%%%%%%%%%%%%%%%%%%%%%%%%%%%%%%
The real root of Eq.(\ref{cubic}) leads to the following,
\begin{eqnarray}
\frac{R}{R_0}~=~\sqrt{\frac{1}{\alpha^2}-1}+{\mathcal{O}}(\beta)\,,
\end{eqnarray}
To make the dependence on the brane world parameter $l_s$ explicit,
we can also write the equation above as,
\begin{eqnarray}
\frac{R}{R_0}~=~\sqrt{\frac{1}{\alpha^2}-1}
+{\mathcal{O}}(\frac{l_{s}^{2}}{R_{0}^{2}})
\nonumber
\end{eqnarray}
These equations are essentially same as Eq.(4)
because corrections are insignificant.
Note that Eq.(7), is a cubic equation, and therefore, there are
two more solutions of the equation. Up to order $\beta$, these two
solutions are coincident, and are given by
\begin{eqnarray}
\frac{R}{R_0}~= {\mathcal{O}}(\beta)
={\mathcal{O}}(\frac{l_{s}^{2}}{R_{0}^{2}})
\nonumber
\end{eqnarray}
It is shown, in appendix-II, that these solutions correspond to
the maxima of the total energy as a function of $R$. They clearly
do not lead to any stable configuration of the compact stars.
As shown in appendix-II, the solution given by Eq.(8), corresponds
to the minimum of the total energy as a function of $R$.
Therefore, in the effective four dimensional theory, the standard
parameters of the white dwarf and neutron stars do not depend on
the brane world induced correction term.However, we note that the
brane world corrections give rise to surface tension which is absent
in Newtonian gravity as well as general theory of relativity.
This surface tension is the new element in gravity. However, we
still do not know  whether it has any observable signature.
\section{Conclusions}
%%%%%%%%%%%%%%%%%%%%%%%%%%%
In this paper we have examined the influence of brane world corrections
on the parameters of white dwarfs and neutron stars using manybody theory.
The brane corrections to Newtonian potential give
rise to interesting new effects, however,
these corrections do not influence the stability of the
stars in a significant way. For instance, the mass,
radius and the mass-radius relation remains
unchanged. We have demonstrated these features
using a simple analytical treatment. Our results
are in qualitative agreement with the numerical work by Wiseman \cite{wise}.

%%%%%%%%%%%%%%%%%%%%%%%%%%%%%%%%%%%%%%%%%%%%%%%%%%%%%%%%%%%%%%
We also point out many new physical features that emerge
from corrections to Newtonian gravity.
These physical features are generic
due to the deviation from the inverse square law of force
and does not depend on the particular form of the correction term:
(a.) In the many body context, force as well as potential due
to the brane correction term depends on density rather than the mass.
Therefore, we expect that the relative
strength of the force compared to Newtonian force will be more
close to the surface of spherical grains of smaller radius.
(b.) Newton-Birkhoff theorem is violated by the
brane world induced correction terms.
Therefore, inside a hollow spherical shell there will be force.
This may turn out to be useful for
experimental verification of Randall-Sundrum or any other similar
corrections to gravity.
%(c.) The General Theory of Relativity (GTR)
%in weak field limit gives rise to  correction terms similar
%to brane corrections which, often, gives rise to the confusion that, on the
%brane, weak field expansion terms due to GTR are similar to higher
%dimensional brane-world induced corrections.
%In the context of GTR, the Newton-Birkhoff theorem remains valid.
%On the other hand, we know that, for any deviation
%of the potential from the Newtonian form,
%there will be shell effect (see the appendix for details). Therefore,
%in GTR when one undertakes weak field expansion,
%one needs to consider successive higher order terms of the
%weak field expansion series to kill this effect. To remove this
%effect completely, one would need to consider the whole expansion series.
(c) Violation of Newton-Birkhoff theorem due to the presence
of brane-world correction terms, gives rise to force similar
to surface tension in normal liquids.Neither Newtonian gravity
nor the general theory of relativity
give rise to force akin to surface tension.
The correction term shares many similarities with
molecular force in normal Liquids. At present,
we are not very clear which of the new features discussed here
are relevant to observations. In our opinion,
this is an important issue which requires
further investigation.

\section{Acknowledgments}
  We thank Roy Maartens for his critical comments on the first draft of the
  manuscript. We also thank N. Dadhich and A. Kembhavi for helpful discussions.

\section{APPENDIX-I}
We want to calculate the gravitational potential energy of a spherically
symmetric ball of radius $R$ and constant density $\rho$.Here the two-body
gravitational potential energy consists of two parts.First part is the standard
Newtonian part given by,
\begin{eqnarray}
U_N=-\frac{Gm_1m_2}{|\vec{r}-\vec{r'}|}
\end{eqnarray}
where $\vec{r}$ and $\vec{r'}$  are the locations of the point masses $m_1$
and $m_2$ respectively.
The Newtonian potential at a point $\vec{r}$ due to a point mass $m$ at point
$\vec{r'}$ is given by
\begin{eqnarray}
\Phi_N(\vec{r})=-\frac{Gm}{|\vec{r}-\vec{r'}|}
\end{eqnarray}

The second part of the potential energy comes
from the brane world correction to gravity which
in the Randall-Sundrum scenario is given by,
\begin{eqnarray}
U_B=-\frac{Gl_{s}^2m_1m_2}{|\vec{r}-\vec{r'}|^3}
\end{eqnarray}
The brane world correction to the potential at point $\vec{r}$ due to
a point mass $m$ at the point $\vec{r'}$ is
\begin{eqnarray}
\Phi_B(\vec{r})=-\frac{Gl_{s}^2m}{|\vec{r}-\vec{r'}|^3}
\end{eqnarray}
For a mass distribution with density $\rho(\vec{r'})$, the formulas above take
the following forms,
\begin{eqnarray}
\Phi_N(\vec{r})=-G\int\frac{\rho(\vec{r'})d^3\vec{r'}}{|\vec{r}-\vec{r'}|}
\end{eqnarray}
\begin{eqnarray}
\Phi_B(\vec{r})=-Gl_{s}^{2}\int\frac{\rho(\vec{r'})d^3\vec{r'}}
{|\vec{r}-\vec{r'}|^3}
\end{eqnarray}
\begin{eqnarray}
U_N=-G\int\int\frac{\rho(\vec{r})\rho(\vec{r'})d^3\vec{r}
d^3\vec{r'}}{|\vec{r}-\vec{r'}|}
=\int\Phi_N(\vec{r})\rho(\vec{r})d^3\vec{r}
\end{eqnarray}
\begin{eqnarray}
U_B=-Gl_{s}^{2}\int\int\frac{\rho(\vec{r})\rho(\vec{r'})d^3\vec{r}
d^3\vec{r'}}{|\vec{r}-\vec{r'}|^3}
=\int\Phi_B(\vec{r})\rho(\vec{r})d^3\vec{r}
\end{eqnarray}

\subsection{Newtonian Potential and Potential Energy for a spherically
symmetric ball of constant density}

Let us consider a spherically symmetric ball  of radius $R$ and  of constant
density $\rho$.In spherical co-ordinate system the formula for $\Phi_{N}(r)$
gives,
\begin{eqnarray}
\Phi_{N}(r)=-G\rho\int\int\int\frac{r'^2dr'sin\theta d\theta d\phi}
{(r^2+r'^2-2rr'cos\theta)^{1/2}}
\end{eqnarray}
In the expression above, the integral over $\phi$ ranges from $0$ to
$2\pi$ and for $\theta$, it ranges from $0$ to $\pi$. The integral over
$\phi$ gives a factor of $2\pi$.
The integral over $\theta$ can be carried out by making the following change
of variable
\begin{eqnarray}
\xi=r^2+r'^2-2rr'cos\theta ;~ d\xi=2rr'sin\theta d\theta~; \nonumber
\end{eqnarray}

The integration over $~\xi~$ ranges from
$~~\xi=(r-r')^2~ ~to ~\xi=(r+r')^2~~ $
After carrying out integrations over $\phi$ and $\theta$, we obtain
\begin{eqnarray}
\Phi_N(r)=~-2\pi G\rho\int[(r+r')-|r-r'|]\frac{r'}{r}dr'
\end{eqnarray}
For $r>~a~(> r')$, and $0\leq r'\leq a$
\begin{eqnarray}
\Phi_N(r)=~-2\pi G\rho\int_{0}^{a}[(r+r')-(r-r')]\frac{r'}{r}dr'=
 -4\pi \frac{G\rho}{r}\int_{0}^{a} r'^2 dr'=-\frac{4\pi a^3\rho G}{3r}=
-\frac{GM}{r}
\end{eqnarray}
For $r~< b~(<r') $, and $b\leq r'\leq R$
\begin{eqnarray}
\Phi_N(r)=~-2\pi G\rho\int_{0}^{a}[(r+r')-(r'-r)]\frac{r'}{r}dr'=
 -4\pi G\rho\int_{b}^{R} r'dr'=-2\pi G\rho(R^2-b^2)
\end{eqnarray}
which is constant independent of $r~$.
Since force is given by the gradient of the potential with respect to $r$,
$F=-grad_{r}~[\Phi_N (r)]$, there is no
gravitational force within the spherical shell.
Once we know the potential , it is easy to calculate the potential energy.
It is given by,
\begin{eqnarray}
U_N=-G\int\int\frac{\rho(\vec{r})\rho(\vec{r'})d^3\vec{r}
d^3\vec{r'}}{|\vec{r}-\vec{r'}|}
=\int\Phi_N(\vec{r})\rho(\vec{r})d^3\vec{r}                \nonumber
\end{eqnarray}
Let us consider the case when $r>r'~$. Taking
$a=r-\epsilon$, we obtain
\begin{eqnarray}
\Phi_N(r)=-\frac{4\pi G\rho (r-\epsilon)^3}{3r}    \nonumber
\end{eqnarray}
The potential energy is given by,
\begin{eqnarray}
U_N= =\int_{(|\vec{r}|\leq R)} \Phi_N(\vec{r})\rho(\vec{r})d^3\vec{r} \nonumber
\end{eqnarray}
Substituting the expression for $\Phi_N(r)$, we obtain,
\begin{eqnarray}
U_N =-\int_{\epsilon}^{R} \frac{4\pi
G\rho (r-\epsilon)^3}{3r}\times 4\pi\rho r^2 dr    \nonumber
\end{eqnarray}
This integral above is convergent, and therefore, we can take the limit
$\epsilon\rightarrow 0~$ before evaluating the integral.After integrating
the resulting expression, we obtain
\begin{eqnarray}
U_N =-\frac{16\pi^2G\rho^2 R^5}{15}=-\frac{3}{5}\times \frac{GM^2}{R}
\end{eqnarray}

\subsection{Potential and Potential Energy for a spherically
symmetric ball of constant density due to Randall-Sundrum Brane world
correction to Gravity}

The potential and the potential energy due to the brane world correction term
is calculated exactly in the same way as in the case of
Newtonian gravity.However, in this case there is a new element.
The outer shell exerts force on the inner shells, and therefore, the effect
of outer shell on the inner shell can not be neglected.

\subsubsection{The Potential at a distance from a spherically  symmetric ball}

\par The potential $\Phi_B(r)$ in spherical co-ordinate
system is given by,
\begin{eqnarray}
\Phi_{B}(r)=-Gl_{s}^{2}\rho\int\int\int\frac{r'^2dr'sin\theta d\theta d\phi}
{(r^2+r'^2-2rr'cos\theta)^{3/2}}
\end{eqnarray}
As in the previous section,
the integral over $\phi$ ranges from $0$ to
$2\pi$ and the integral over $\theta$ ranges from $0$ to $\pi$.
Integral over
$\phi$ gives factor of $2\pi$.
Integral over $\theta$ can be carried out by making the following change
of variable
\begin{eqnarray}
\xi=r^2+r'^2-2rr'cos\theta ;~~ d\xi=2rr'sin\theta d\theta~; \nonumber
\end{eqnarray}

The integration over $~\xi~$ ranges from
$~~\xi=(r-r')^2~ ~to~ ~\xi=(r+r')^2~~ $
After carrying out integrations over $\phi$ and $\theta$, we obtain
\begin{eqnarray}
\Phi_B(r)=~-2\pi Gl_{s}^{2}\rho\int[\frac{1}{|r-r'|}-\frac{1}{r+r'}]
\frac{r'}{r}dr'
\end{eqnarray}
For $r>a~(>r')$, and $0\leq r'\leq a$
\begin{eqnarray}
\Phi_B(r)=~-2\pi Gl_{s}^{2}\rho\int_{0}^{a}[\frac{1}{r-r'}-\frac{1}{r+r'}]
\frac{r'}{r}dr'
\end{eqnarray}
After some algebraic manipulations, this integral can be evaluated.The
result is
\begin{eqnarray}
\Phi_B(r)=~-2\pi G~l_{s}^{2}~\rho~[~ln\frac{r+a}{r-a}~-~\frac{2a}{r}~]
\end{eqnarray}
Large distance limit of the potential can be evaluated in two different ways.
From the formula,
\begin{eqnarray}
\Phi_B(\vec{r})=-Gl_{s}^{2}\int_{0}^{a}\frac{\rho(\vec{r'})d^3\vec{r'}}
{|\vec{r}-\vec{r'}|^3}                                  \nonumber
\end{eqnarray}
we find that for $r>>a$,
\begin{eqnarray}
\Phi_B(\vec{r})=-\frac{Gl_{s}^{2}}{|\vec{r}|^3}
\int_{0}^{a}\rho(\vec{r'})d^3\vec{r'}
=-\frac{4\pi Gl_{s}^{2}\rho a^3}{3r^3}=-\frac{Gl_{s}^{2}M}{r^3}   \nonumber
\end{eqnarray}
The other way is to consider the formula,
\begin{eqnarray}
\Phi_B(r)=~-2\pi G~l_{s}^{2}~\rho~[~ln\frac{r+a}{r-a}~-~\frac{2a}{r}~]
                                                 \nonumber
\end{eqnarray}
which has been derived for $r>a$, and therefore, can also be written as,
\begin{eqnarray}
 \Phi_B(r)=~-2\pi G~l_{s}^{2}~\rho~[~ln\frac{1+a/r}{1-a/r}~-
~\frac{2a}{r}~]
=~-2\pi G~l_{s}^{2}~\rho~[~ln(1+a/r)-ln(1-a/r)~-~\frac{2a}{r}~]
                     \nonumber
\end{eqnarray}
Replacing the logarithms above by their series expansion, after some
algebraic manipulations we obtain,
\begin{eqnarray}
\Phi_B(r)=-\frac{4\pi Gl_{s}^{2}\rho a^3}{3r^3}
-\frac{4\pi Gl_{s}^{2}\rho a^5}{5r^5}
-\frac{4\pi Gl_{s}^{2}\rho a^7}{7r^7}-....
=-\frac{Gl_{s}^{2}M}{r^3}-... ... ...
                      \nonumber
\end{eqnarray}

\subsubsection{Force at a distance from a spherically symmetric ball}

Force at a distance from a spherically symmetric ball can be calculated
by evaluating the gradient of the potential,
$F=-Grad\Phi_B(r)=-\frac{d\Phi_B(r)}{dr}$.
The potential is given by,
\begin{eqnarray}
\Phi_B(r)=~-2\pi G~l_{s}^{2}~\rho~[~ln\frac{r+a}{r-a}~-~\frac{2a}{r}~]
                                                 \nonumber
\end{eqnarray}
and therefore, the force is given by
\begin{eqnarray}
F=2\pi Gl_{s}^{2}~\rho~[\frac{1}{r+a}-\frac{1}{r-a}+\frac{2a}{r^2}]
=-4\pi aGl_{s}^{2}~\rho~[\frac{1}{r^2-a^2}-\frac{1}{r^2}]
\end{eqnarray}
Let us calculate the force at a very small distance from the surface.
We take $r-a=\epsilon$, $~\epsilon$ very small, and substitute it in the
equation above.After some algebraic manipulation we obtain,
\begin{eqnarray}
F=2\pi Gl_{s}^{2}\rho~[-\frac{1}{\epsilon}+\frac{2}{a}]
\end{eqnarray}
This is clearly a skin effect. However, this force is not as strong as one may
think. At a distance of $1~mm$ from the surface of a neutron star, this force
is weaker than the force of Newtonian gravity.The reason is that this force
depends on the density and not the mass. However, this force can be large
close to the surface of a small spherical grain and may even overtake
the force due to Newtonian gravity.

\subsubsection{Potential inside a Spherical Shell}

Let us consider a spherical shell of outer
radius $R$ and inner radius $b$. At a point $r$ inside the shell,
the potential is
\begin{eqnarray}
\Phi'_B(r)=~-2\pi Gl_{s}^{2}\rho\int_{b}^{R}[\frac{1}{|r-r'|}-\frac{1}{r+r'}]
\frac{r'}{r}dr'
                        \nonumber
\end{eqnarray}
Since $r<b\leq r'\leq R$, the formula above becomes,
\begin{eqnarray}
 \Phi'_B(r)=-2\pi Gl_{s}^{2}\rho\int_{b}^{R}[\frac{1}{r'-r}-\frac{1}{r+r'}]
\frac{r'}{r}dr'                     \
=-2\pi Gl_{s}^{2}\rho\int_{b}^{R}\frac{2r'd'}{r'^2-r^2}
=-2\pi Gl_{s}^{2}\rho~ ln\frac{R^2-r^2}{b^2-r^2}
\end{eqnarray}
\subsubsection{Force inside the shell}
Force inside the shell can be calculated by taking the gradient of the
potential, $F=-Grad~[\Phi'_B(r)]=-\frac{d\Phi'_B(r)}{dr}$.
\par For the potential,
\begin{eqnarray}
\Phi'_B(r)=-2\pi Gl_{s}^{2}\rho~ ln\frac{R^2-r^2}{b^2-r^2} \nonumber
\end{eqnarray}
we obtain,
\begin{eqnarray}
F(r)=2\pi Gl_{s}^{2}\rho~[\frac{2r}{b^2-r^2}-\frac{2r}{R^2-r^2}]
\end{eqnarray}
From the expression above it is obvious that $~~F(r)>0$.
This means that it is directed in the direction in which
$r$ increases.In other words, it is directed towards the
inner boundary surface of the  shell.
At the centre of the shell, $r=0$, and therefore,
the force is zero.But any small perturbation will move the test particle
away towards the inner boundary surface of the shell.
This is a new element that emerges
from the brane world correction to Newton's law of gravity. We know that the
force due to Newtonian gravity
is zero inside a spherically symmetric shell. If one can find the above
mentioned effect inside a shell, it will provide a very  clean test of
brane world corrections to Newtonian gravity. May be such experiments can be
carried out on board artificial satellites.
\par In this case, as in the case of spherical ball, there is skin effect.
The force of attraction close to the inner
boundary of the shell becomes large.
This can be seen as follows. Let $b=r+\epsilon~~$.
This simply means that the layer
at $r$ is just bellow the inner boundary of the shell
($\epsilon$ distance away from the inner surface).
\par In this case the force at $~r~$ is given by,
\begin{eqnarray}
F=2\pi Gl_{s}^{2}\rho~[\frac{1}{\epsilon}-\frac{2r}{R^2-r^2}]
\end{eqnarray}

\subsubsection{Cancellation of the Skin Effect on the Inner Layers of
a spherical solid ball}

Any interior layer of the spherical ball experiences two types
of forces:~
one is due to a sphere just bellow it and other is due to
the spherical shell outside.
These  two types of forces give rise to a large force on this layer
due to the skin effect.
Remarkably, these two forces act in opposite directions and the forces due
to skin effect are equal in magnitude and exactly cancel each other.It is easy
to verify this.
On the layer at $r$, let $F_1$ be the force due the inner sphere and $F_2$
be the force due to the outer shell.
\begin{eqnarray}
F_1=2\pi Gl_{s}^{2}\rho~[-\frac{1}{\epsilon}+\frac{2}{r}]
\end{eqnarray}
\begin{eqnarray}
F_2=2\pi Gl_{s}^{2}\rho~[\frac{1}{\epsilon}-\frac{2r}{R^2-r^2}]
\end{eqnarray}
Therefore, the net force at any point on a layer at $r$, is given by
\begin{eqnarray}
F=F_1+F_2=2\pi Gl_{s}^{2}\rho~[\frac{2}{r}-\frac{2r}{R^2-r^2}]
\end{eqnarray}
This result is valid for every layer
except the one at the outer most boundary.
For the outer most layer, there is an inner sphere but no outer shell.
There is no cancellation of the skin effect on this layer.
Therefore, it experiences a large force. This will give rise to effect
similar to surface tension in normal liquids.

\subsubsection{The Potential Energy}

We found in the previous sections that, for a  sphere ball,
the potential on a given layer at a radial distance $r$
from the centre consists of two parts.First part is due to a
interior spherical ball. It is given by,
\begin{eqnarray}
\Phi_B(r)=~-2\pi G~l_{s}^{2}~\rho~[~ln\frac{r+a}{r-a}~-~\frac{2a}{r}~]
                                                 \nonumber
\end{eqnarray}
The second part is due to the outer spherical shell. It is given by,
\begin{eqnarray}
\Phi'_B(r)=~-2\pi Gl_{s}^{2}\rho~ ln\frac{R^2-r^2}{b^2-r^2} \nonumber
\end{eqnarray}
The potential energy also consists of two parts.
The first part of the potential energy is obtained by taking the radius
of inner sphere $a=r-\epsilon$ in the formula for the potential,
multiplying the potential by $4\pi r^2 dr\rho$, the mass of the thin layer
and then integrating the resulting expression from $r=\epsilon$
to $r=R$, where $\epsilon$ is the smallest inter particle distance,
$R$ is the total radius of the spherical ball.
This is given by,
\begin{eqnarray}
U_1=~-2\pi Gl_{s}^{2}\rho~\int_{\epsilon}^{R}
[ln\frac{2r-\epsilon}{\epsilon}-\frac{2(r-\epsilon)}{r}]4\pi\rho r^2dr
\end{eqnarray}
The second part of the potential energy is obtained by taking
the inner radius of the spherical shell, $b=r+\epsilon$
in the formula for the potential due to the shell,
multiplying the potential by $4\pi r^2 dr\rho$, the mass of the thin
layer, and then integrating the resulting expression from $r=0$
to $r=R$, where $R$ is the total radius of the spherical ball.
This is given by,
\begin{eqnarray}
U_2=~-2\pi Gl_{s}^{2}\rho~\int_{0}^{R}
[ln\frac{R^2-r^2}{(r+\epsilon)^2-r^2}]4\pi\rho r^2dr
\end{eqnarray}
After some algebraic manipulations, we obtain
\begin{eqnarray}
&& U_2=~-2\pi Gl_{s}^{2}\rho~\int_{0}^{R}
[ln\frac{R+r}{\epsilon}+ln\frac{R-r}{\epsilon}\nonumber \\
&&-ln\frac{2r+\epsilon}{\epsilon}]4\pi\rho r^2dr      \nonumber
\end{eqnarray}
The total potential energy is given by,
\begin{eqnarray}
U&=&U_1+U_2  \nonumber\\
&=&-8\pi^2Gl_{s}^{2}\rho^2~[\int_{\epsilon}^{R}r^2dr
(ln\frac{2r-\epsilon}{\epsilon}-\frac{2(r-\epsilon)}{r}) \nonumber \\
&& +\int_{0}^{R}r^2dr(ln\frac{R+r}{\epsilon}+\nonumber \\
&&ln\frac{R-r}{\epsilon}
-ln\frac{2r+\epsilon}{\epsilon})]   \nonumber
\end{eqnarray}
:In the $\epsilon\rightarrow~0$ limit, it is not difficult to show that
\begin{eqnarray}
\int_{\epsilon}^{R}ln\frac{2r-\epsilon}{\epsilon}r^2dr
-\int_{0}^{R}ln\frac{2r+\epsilon}{\epsilon}r^2dr~~~ \rightarrow 0
\end{eqnarray}
Therefore, the potential energy is given by,
\begin{eqnarray}
U=~-8\pi^2Gl_{s}^{2}\rho^2~[\int_{\epsilon}^{R}r^2dr
(-\frac{2(r-\epsilon)}{r})
+\int_{0}^{R}r^2dr(ln\frac{R+r}{\epsilon}+ln\frac{R-r}{\epsilon})]
\end{eqnarray}
The last two integrals after change of variables amounts to evaluating
standard table integrals of the form $\int x^nlnx$.We simply write down
the results.
\begin{eqnarray}
\int_{0}^{R}ln\frac{R+r}{\epsilon} r^2dr=\frac{1}{3}R^3ln~\frac{R}{\epsilon}
+\frac{2}{3}~R^3~ln2-\frac{5}{18}~R^3        \nonumber
\end{eqnarray}
\begin{eqnarray}
\int_{0}^{R}ln\frac{R-r}{\epsilon} r^2dr=\frac{1}{3}R^3ln~\frac{R}{\epsilon}
-\frac{11}{18}~R^3 \nonumber
\end{eqnarray}
\begin{eqnarray}
\int_{\epsilon}^{R}r^2dr (-\frac{2(r-\epsilon)}{r})=-\frac{2}{3}R^3
                                                            \nonumber
\end{eqnarray}
Therefore,
\begin{eqnarray}
U&=&-8\pi^2Gl_{s}^{2}\rho^2~[~~\frac{2}{3}~R^3~ln~\frac{R}{\epsilon}
+\frac{2}{3}~R^3~ln~2~-~\frac{14}{9}~R^3~~]
=-8\pi^2Gl_{s}^{2}\rho^2~[~~\frac{2}{3}~R^3~ln~\frac{R}{\epsilon}
-\frac{2}{3}~R^3~ln~(5.16)~~] \nonumber\\
&=&-8\pi^2Gl_{s}^{2}\rho^2~[~~\frac{2}{3}~R^3~ln(~\frac{R}{5.16\epsilon})]
                                                                \nonumber
\end{eqnarray}
With a redefined $\epsilon$, we can write it as,
\begin{eqnarray}
U= -\frac{16\pi^2}{3}Gl_{s}^{2}\rho^2~R^3~ln\frac{R}{\epsilon}
=-\frac{3GM^2l_{s}^{2}}{R^3}ln\frac{R}{\epsilon}               \nonumber\\
\end{eqnarray}
If $N$ is the number of particles in the spherical ball and $m$ is the
mass of the elementary constituent particles, then $M=Nm~$ and we can
write the total potential energy due to Randall-Sundrum brane-world
induced correction term as
\begin{eqnarray}
U=U_{RS}= -\frac{3GN^2 m^2 l_{s}^{2}}{R^3}ln\frac{R}{\epsilon}
\end{eqnarray}
The small parameter, $\epsilon$, in the formulae above should be
interpreted as the minimum inter-particle distance in the spherically
symmetric volume.

\section{APPENDIX-II}
We want to calculate the extrema of the energy functionals considered
in the main text of the paper. We will consider only the white dwarf
stars. The results for neutron stars can be obtained simply by replacing
the electron mass by the neutron mass.\\
Most literature on white dwarf and neutron stars treat the
ultra-relativistic limit of such stars taking the kinetic energy term as,
$K.E.=Nc|p|$. We will first show that in the presence of brane corrections
such systems do not have any stable quantum mechanical ground state.
In this case the total energy including the
brane world corrections as a function of $R$,  is given by
\begin{eqnarray}
E(R)=N~c~|p|-\frac{3}{5}\times \frac{Gm_{n}^{2} N^2}{R}
-\frac{3Gm_{n}^{2} N^2 l_{s}^{2}}{R^3}\times ln\frac{R}{\epsilon}
\nonumber
\end{eqnarray}
Substituting $p=\frac{\hbar N^{1/3}}{R}$, we obtain
\begin{eqnarray}
E(R)=\frac{c\hbar N^{4/3}}{R}-\frac{3}{5}\times \frac{Gm_{n}^{2} N^2}{R}
-\frac{3Gm_{n}^{2} N^2 l_{s}^{2}}{R^3}\times ln\frac{R}{\epsilon}
\nonumber
\end{eqnarray}
With $N_{max}$ and $\alpha$, defined as
\begin{eqnarray}
N_{max}=\frac{5}{3}\times \frac{c\hbar}{Gm_{n}^{2}}  ~~~;
~~~\alpha=(N/N_{max})^{2/3}
\nonumber
\end{eqnarray}
we can write the total energy given above as
\begin{eqnarray}
E(R)=c\hbar N^{4/3}[\frac{1}{R}-\frac{\alpha}{R}-\frac{15\alpha l_{s}^{2}}
{R^3}\times ln\frac{R}{\epsilon}]
\nonumber
\end{eqnarray}
The extremum is obtained by equating the the derivative of $E(R)$ with
respect to $R$ to zero with $N$ held fixed.
\begin{eqnarray}
\frac{dE}{dR}=-\frac{c\hbar N^{4/3}}{R^2}(1-\alpha-\frac{45\alpha l_{s}^{2}}
{R^2}\times ln\frac{R}{\epsilon_1})=0
\nonumber
\end{eqnarray}
where $\epsilon_1=\gamma \epsilon~~;~~~\gamma~$ being some positive
number,$~1<\gamma<10~$.
To find the nature of the extremum, we calculate the second derivative
of $E(R)$ with respect to $R$.
\begin{eqnarray}
\frac{d^2E}{dR^2}=
\frac{2c\hbar N^{4/3}}{R^3}(1-\alpha-\frac{45\alpha l_{s}^{2}}
{R^2}\times ln\frac{R}{\epsilon_1})
-\frac{c\hbar N^{4/3}}{R^2}(\frac{90~\alpha~ l_{s}^{2}}{R^3}\times
ln\frac{R}{\epsilon_2})
\nonumber
\end{eqnarray}
where $\epsilon_2=\gamma_1 \epsilon_1~~;~~~\gamma_1~$ being some positive
number, $~1<\gamma_1<10~$. The quantities $\epsilon$, $\epsilon_1$
and $\epsilon_2$ as an order of magnitude can be taken to be
equal to $\frac{R}{N^{1/3}}$.
In the equation above, the first term is zero from the condition that
$\frac{dE}{dR} = 0$. The second term is strictly negative. Therefore, the
extremum of the function, $E(R)$, is a maximum.
The energy $E(R)$ as a function
of $R$, starts from $-\infty$ at $R=0$, rises very sharply
to a positive value and develops a maximum very close to $R=0$,
and then from the positive side of energy it
slowly approaches zero at $R=\infty$.

\par Next we consider the total energy functional given by
\begin{eqnarray}
E(R)= K_N~+~U_{N}(R)
= N~c~(p^2+m_{e}^{2}c^2)^{1/2} -Nm_e c^2
-\frac{3}{5}\times \frac{Gm_{n}^{2} N^2}{R}
                                                         \nonumber
\end{eqnarray}
Substituting $p=\frac{\hbar N^{1/3}}{R}$, we obtain
\begin{eqnarray}
E(R) = \nonumber N~c~(\frac{\hbar^2 N^{2/3}}{R^2}+m_{e}^{2}c^2)^{1/2}
-Nm_e c^2 -\frac{3}{5}\times \frac{Gm_{n}^{2} N^2}{R}
                                                   \nonumber
\end{eqnarray}
The extremum with respect to $R$ is given by the condition,
\begin{eqnarray}
\frac{dE}{dR}=-\frac{c\hbar N^{4/3}}{R^2}\Big{[}
\frac{N^{1/3}\hbar}{(\hbar^2 N^{2/3}+m_{e}^{2}c^2 R^2)^{1/2}}
-\frac{3}{5}\times \frac{Gm_{n}^{2} N^{2/3}}{c \hbar}\Big{]}=0
\nonumber
\end{eqnarray}
In terms of the parameters, $R_0$ ($~=\frac{\hbar N^{1/3}}{m_e c}$) and
$\alpha$ , the equation above simplifies to the form
\begin{eqnarray}
\frac{dE}{dR}=-\frac{c\hbar N^{4/3}}{R^2}
\Big{[}\frac{1}{(1+\frac{R^2}{R_{0}^{2}})^{1/2}}-\alpha\Big{]}=0
\nonumber
\end{eqnarray}
To find the nature of the extremum, we calculate the second derivative
of $E(R)$ with respect to $R$. It is given by,
\begin{eqnarray}
\frac{d^2E}{dR^2}=\frac{2c\hbar N^{4/3}}{R^3}
\Big{[}\frac{1}{(1+\frac{R^2}{R_{0}^{2}})^{1/2}}-\alpha\Big{]}\,
-\frac{c\hbar N^{4/3}}{R^2}
\Big{[}\frac{-R}{R_{0}^{2}(1+\frac{R^2}{R_{0}^{2}})^{3/2}}\Big{]}
\nonumber
\end{eqnarray}
The first term is zero because of the condition, $\frac{dE}{dR}=0$.
Therefore,
\begin{eqnarray}
\frac{d^2E}{dR^2}= \frac{c\hbar N^{4/3}}{R}
\Big{[}\frac{1}{R_{0}^{2}(1+\frac{R^2}{R_{0}^{2}})^{3/2}}\Big{]}
\nonumber
\end{eqnarray}
which is strictly positive implying that the extremum is a minimum.
The energy $E(R)$ as a function of $R$, starts from $\infty$ at $R=0$
and rolls down to a minimum at a negative value of $E(R)$ and then
from the negative side of energy it slowly approaches zero at $R=\infty$.

\par Now we consider extremum of the total energy functional given by,
\begin{eqnarray}
E(R)= K_N~+~U_{N}(R)+~U_{RS}
= N~c~(p^2+m_{e}^{2}c^2)^{1/2} -Nm_e c^2
-\frac{3}{5}\times \frac{Gm_{n}^{2} N^2}{R}
-\frac{3Gls^{2}m_{n}^{2}N^2}{R^3}\times ln\frac{R}{\epsilon}
                                                         \nonumber
\end{eqnarray}
Substituting $p=\frac{\hbar N^{1/3}}{R}$, we obtain
\begin{eqnarray}
E(R) = N~c~(\frac{\hbar^2 N^{2/3}}{R^2}+m_{e}^{2}c^2)^{1/2} -Nm_e c^2
-\frac{3}{5}\times \frac{Gm_{n}^{2} N^2}{R}
-\frac{3Gls^{2}m_{n}^{2}N^2}{R^3}\times ln\frac{R}{\epsilon}
                                                   \nonumber
\end{eqnarray}

The necessary condition for the  extremum is given by,
\begin{eqnarray}
\frac{dE(R)}{dR}=-\frac{c~ N^{5/3} \hbar^2}{R^3
(\frac{\hbar^2 N^{2/3}}{R^2}+m_{e}^{2}c^2)^{1/2}}
+\frac{3}{5}\times \frac{Gm_{n}^{2} N^2}{R^2}
+\frac{9Gls^{2}m_{n}^{2}N^2}{R^4}\times ln\frac{R}{\epsilon}
-\frac{3Gls^{2}m_{n}^{2}N^2}{R^4}=0      \nonumber
\end{eqnarray}
The last term, in the equation above, can be absorbed in the third term
containing the logarithm by simply redefining the parameter $\epsilon$.
As before, we denote it by $\epsilon_1$.
With the redefined $\epsilon$, the equation above takes the
following form,
\begin{eqnarray}
\frac{dE}{dR}=-\frac{c\hbar N^{4/3}}{R^2}\Big{[}
\frac{N^{1/3}\hbar}{(\hbar^2 N^{2/3}+m_{e}^{2}c^2 R^2)^{1/2}}
-\frac{3}{5}\times \frac{Gm_{n}^{2} N^{2/3}}{c \hbar}
-\frac{9Gls^{2}m_{n}^{2}N^{2/3}}{R^2 c\hbar}\times ln\frac{R}{\epsilon_1}
\Big{]}=0
\nonumber
\end{eqnarray}
In terms of the parameters, $R_0$ and
$\alpha$ , the equation above simplifies to the form
\begin{eqnarray}
\frac{dE}{dR}=-\frac{c\hbar N^{4/3}}{R^2}
\Big{[}\frac{1}{(1+\frac{R^2}{R_{0}^{2}})^{1/2}}-\alpha
-\frac{15 \alpha~l_{s}^{2}}{R^2}\times ln\frac{R}{\epsilon_1}\Big{]}=0
\nonumber
\end{eqnarray}
In the main text of the paper, it is shown that
\begin{eqnarray}
\frac{R^2}{R_{0}^{2}}= \frac{1}{\alpha^2}-1+{\mathcal O}(\beta)
\nonumber
\end{eqnarray}
solves the equation above.
The parameter $\beta$ is defined in the text as,
\begin{eqnarray}
\beta=\frac{15 l_{s}^{2}}{R_{0}^{2}}\times ln\frac{R}{\epsilon}
\approx \frac{15 l_{s}^{2}}{R_{0}^{2}}\times ln N^{1/3}
\nonumber
\end{eqnarray}
For $l_{s}=1mm$ and $N\approx 10^{57}$, we find that
$\beta \approx \frac{3}{R_{0}^{2}} $.
Therefore, the solution of the equation for the extremum can also
be written as
\begin{eqnarray}
\frac{R^2}{R_{0}^{2}}= \frac{1}{\alpha^2}-1+{\mathcal O}(\frac{1}{R_{0}^{2}})
\end{eqnarray}
To make the brane world parameter dependence explicit, we can also write
it as
\begin{eqnarray}
\frac{R^2}{R_{0}^{2}}= \frac{1}{\alpha^2}-1+{\mathcal O}
(\frac{l_{s}^{2}}{R_{0}^{2}})
\end{eqnarray}

To find the nature of the extremum, we need to calculate the second derivative
with respect to $R$ of the function $E(R)$. This is given by,
\begin{eqnarray}
\frac{d^2E}{dR^2}=\frac{2c\hbar N^{4/3}}{R^3}
\Big{[}\frac{1}{(1+\frac{R^2}{R_{0}^{2}})^{1/2}} -\alpha
-\frac{15 \alpha~l_{s}^{2}}{R^2}\times ln\frac{R}{\epsilon_1}\Big{]}
-\frac{c\hbar N^{4/3}}{R^2}
\Big{[}\frac{-R}{R_{0}^{2}(1+\frac{R^2}{R_{0}^{2}})^{3/2}}
+\frac{30 \alpha~l_{s}^{2}}{R^3}\times ln\frac{R}{\epsilon_2}\Big{]}
\nonumber
\end{eqnarray}
The first term in the equation above is zero because of the condition,
$\frac{dE}{dR}=0$. With $l_{s}=1mm=0.1 cm$ and $\epsilon_2=\frac{R}{N^{1/3}}$,
we can write the equation above as,
\begin{eqnarray}
\frac{d^2E}{dR^2}=
\frac{c\hbar N^{4/3}}{R}
\Big{[}\frac{1}{R_{0}^{2}(1+\frac{R^2}{R_{0}^{2}})^{3/2}}
-\frac{0.3 \alpha~ lnN^{1/3}}{R^4}\Big{]}
\nonumber
\end{eqnarray}
Using Eq.( ), this can be written as,
\begin{eqnarray}
\frac{d^2E}{dR^2}=
\frac{c\hbar N^{4/3}\alpha^3}{R}
\Big{[}\frac{1}{R_{0}^{2}+{\mathcal O}(l_{s}^{2})}
-\frac{0.3 \alpha^2~ lnN^{1/3}}{R_{0}^{4}(\alpha^2-1)^2+
{\mathcal O}(l_{s}^{2}R_{0}^{2}) }\Big{]}
\end{eqnarray}
For all admissible values of the parameters for  white dwarfs as well
as neutron stars, the expression above is strictly positive. Therefore,
the extremum is a minimum.
Note that the equation for extrema is a cubic equation in
$R^2$, and therefore, there should be two more solutions. These two
solutions up to order $\beta$ are real and coincident.These two
solutions are given by,
\begin{eqnarray}
\frac{R^2}{R_{0}^{2}}={\mathcal O}(\beta)={\mathcal O}
(\frac{l_{s}^{2}}{R_{0}^{2}})
\end{eqnarray}
These solutions correspond to a maximum of the total energy $E(R)$.
The energy $E(R)$ as a function
of $R$, starts from $-\infty$ at $R=0$, rises very sharply
to a positive value and develops a maximum very close to $R=0$,
it rolls down to a minimum at a negative value of $E(R)$ and then
from the negative side of energy it slowly approaches zero at $R=\infty$.

% \section{Acknowledgments}
%  We thank Roy Maartens for his critical comments on the first draft of the
%manuscript. We also thank N. Dadhich for helpful discussions.

\end{document}